\documentstyle[12pt]{article}
\begin{document}
\title{Screening Current Effects in Josephson Junctions Arrays}
\author{A. Petraglia$^{1,2}$, G.  Filatrella$^{2,3}$, G. Rotoli$^4$}
\date{July 27, 1995\\ PACS \# 74.50.+r, 75.50.LK}
\maketitle
\noindent
1. Physics Department, DTU, Dk-2800 Lyngby, Denmark\\
2. Unit\`a INFM and Dipartimento di Fisica\\
University of Salerno, I-84081 Baronissi (SA), Italy\\
3. Physikalisches Institut, Lehrstuhl Experimentalphysik II, University
of T\"ubingen\\
D-72076 T\"ubingen, Germany\\
4. Unit\`a INFM and Dipartimento di Energetica, University of L'Aquila,
I-67010 Localit\`a Monteluco-Roio Poggio (AQ), Italy

\begin{abstract}

The purpose of this work is to compare the  dynamics of  arrays of
Josephson junctions in presence of magnetic field in two different
frameworks: the so called XY frustrated model with no self inductance
and an approach that takes into account the screening currents
(considering self inductances only). We show that while for a range of
parameters the simpler model is sufficiently accurate, in a region of
the parameter space solutions arise that are not contained in the XY
model equations.

\end{abstract}

\baselineskip=15pt
\section{Introduction}
Arrays of Josephson junctions have been proposed more than two decades
ago to enhance the emission of microwave {\bf \cite{tilley70}}. In fact
it is
well known that the power available from a single junctions is not
enough for many practical applications and therefore the achievement of
coherent motion of arrays of junctions is an important issue for device
applications
{\bf \cite{wiesenfeld94}}. Apart from the applications, two-dimensional
arrays have been also investigated as an interesting nonlinear system
both experimentally and theoretically {\bf
\cite{parment94,dominguez94}}.  In the study of such arrays two classes
of models have been proposed to describe the dynamics:

\begin{itemize}
\item[1)] Models that neglect the effect of the screening currents.
Those model are often called ``Uniformly Frustrated XY''
{\bf \cite{tei,hal,gro}}
because the Hamiltonian for this model is similar to that of a square
flat lattice of spins in magnetic field (for that reason it is also
called Spin Glass model).
\item[2)]Models that include, to some extent, the effect of the
screening
currents. The simplest version of this model assumes that the effect of
the screening is to shield the magnetic field only in the two cells
adjacent to each branch where a current is flowing (in the following
this model will be termed the NS model after Nakajima and Sawada
that have introduced it {\bf \cite{nakajima81}}).
More
generally one should consider the effect of the screening in all cells
(full mutual inductance approach {\bf \cite{phillips93,reinel94}}).
\end{itemize}

To neglect screening currents greatly simplifies the problem.
Moreover, the formal similarity with other well known systems allows to
take advantage of an accumulated experience in those contexts.
Model (2) has been used less extensively
also
because, in its complete form (that includes all mutual inductances),
requires
complex numerical routines and the resulting equations are difficult
to handle analytically {\bf \cite{phillips93,reinel94}}.
In spite of the simplicity of model (1), it has been able to explain
most of the experimental observations on two-dimensional arrays, for
example giant Shapiro steps {\bf \cite{benz90}}.
It is therefore of interest to
establish the limits of validity of this approach, that should be taken
into account when interpreting the experimental data.
Moreover, to develop coherent sources based on large
two-dimensional arrays of Josephson junctions, better performances can
be obtained if the junctions in the array have larger
critical currents; in fact the emitted power of a single junction is
proportional to the square of the critical current, while the emitted
power of an ideal coherent array will $\propto N^2 I_0^2$, where $N$ is
the number of locked junctions in the array {\bf \cite{barone82}}.
This, as will be shown in the following, prevents for practical
purposes to use arrays whose coupling is so strong to avoid large
screening currents and therefore the  model (2) seems most suitable to
describe arrays usable as microwave sources.

The purpose of this work is to explore the parameter space looking for
the limits of validity of model (1).
We will show that, for certain ranges of the parameters and at least
for the simplest arrays, the solutions of model (1) are really
different from the solutions of model (2).
To fully understand the differences between the two approaches we will
present a step by step derivation of the dynamical equations. As a
tutorial example we will start with a discussion of a simple dc SQUID
(Superconducting QUantum Interferometer) containing only two junctions.
This well known case will lead us to some general considerations on
the role of the screening currents.

It will be shown that the XY equations cannot be easily derived from NS
equations taking some limits of the parameters. It will be further shown
numerically that there is a region of the parameter space where it is
not possible to assume that the screening current is negligible. Next,
we will consider the most elementary cell of a two-dimensional array,
{\em i.e.} a square cell with junctions (and inductances) on each branch
{\bf \cite{sohn94}}. We will again derive the equations and we will
numerically show that the screening current are not negligible for some
ranges of the parameters. A comparison with the actual two-dimensional
arrays will be performed for some cases to check that the results for
the single cell are reliable. Finally, the macroscopic dynamics in the
two regimes (with negligible and not negligible screening currents) will
also be shown.

\section{The Squid model}
A dc SQUID is a superconducting loop interrupted by two junctions {\bf
\cite{barone82}}. Since there is a single superconducting loop the flux
quantization, neglecting screening currents, is given by:

%Eq. 1
\begin{equation}
\phi_1 - \phi_2 = 2 \pi f \end{equation}
\noindent where $\phi_1$ and $\phi_2$ are the gauge invariant phases
across the two junctions, $f=\Phi_a/\Phi_0$ is the frustration, or the
ratio between the applied flux $\Phi_a$ and the quantum flux $\Phi_0
=\hbar/2e$. Assuming the RSJ model for the junctions {\bf
\cite{barone82}} the equation of motion of the phase difference is given
by the current balance across a branch. In normalized units it reads:
%Eq.s 2,3
\begin{eqnarray}
\ddot{\phi_1} + \alpha \dot{\phi_1} + \sin \phi_1 &=& \gamma,\\
\ddot{\phi_2} +\alpha  \dot{\phi_2} + \sin \phi_2 &=& \gamma.
\end{eqnarray}
\noindent
Here time is normalized with respect to the
inverse of the Josephson frequency $\omega_j=2\pi I_0/C\Phi_0$ ($C$ is
the capacitance of the junction and $I_0$ is the maximum Josephson
current), $\alpha = (\Phi_0/2\pi C R^2 I_0)^{1/2}$ is the damping term
($R$ is the shunt resistance), $\gamma$ is the bias current normalized
to $I_0$. Here and in the following we will assume the parameters of
the junctions to be identical.\\
Taking the sum of Eq.s (2,3) and
inserting Eq. (1) the equation of motion for $\phi_1$ is readily
obtained:
%Eq. 4
\begin{equation} \gamma = \ddot{\phi_1} + \alpha
\dot{\phi_1} + \frac{1}{2} \left[ \sin \phi_1 + \sin ( \phi_1 - 2 \pi
f)\right] .
\end{equation}
\noindent Under the approximation of
negligible screening current the SQUID is described by Eq. (4).
Although
it is possible to recognize some characteristics of the SQUID, in some
respect Eq. (4) is qualitatively incomplete. To better recognize the
difference it is useful to follow the complete derivation of the
equation considering also the screening current {\bf \cite{tesche77}},
{\em i.e.}, modifying Eq. (1) as follows:
%Eq. 5
\begin{equation}
\phi_1 - \phi_2 = 2 \pi f + \frac{2\pi L}{\Phi_0} I_s \end{equation}
\noindent (here $L$ is the inductance of the superconducting loop and
$I_s$ is the screening current in the loop).  The current balance now
is modified by the presence of the screening current:
%Eq.s 6,7
\begin{eqnarray}
\ddot{\phi_1} + \alpha \dot{\phi_1} + \sin\phi_1 & =&
\gamma - \frac{I_s}{I_0},\\ \ddot{\phi_2} + \alpha \dot{\phi_2} + \sin
\phi_2 & =&
 \gamma + \frac{I_s}{I_0}.  \end{eqnarray}

\noindent Inserting Eq. (5) will lead to two coupled equations rather
than one ($\beta_l$ is the SQUID parameter $\beta_l = 2\pi L
I_0/\Phi_0$)
%Eq.s 8,9
\begin{eqnarray}
\ddot{\phi_1} + \alpha \dot{\phi_1} + \sin \phi_1 &
=&
\gamma - \frac{1}{\beta_l}(\phi_2 - \phi_1) + \frac{2\pi}{\beta_l} f,\\
\ddot{\phi_2} + \alpha \dot{\phi_2} + \sin \phi_2 &
=&
\gamma + \frac{1}{\beta_l}(\phi_2 - \phi_1) - \frac{2\pi}{\beta_l} f.
\end{eqnarray} \noindent

The most obvious difference between the Eq.s (8,9) and the
XY model is the increased number of equations: while in the XY approach
the dynamics of the junctions are essentially identical [apart from an
additive constant, see Eq. (1)] in the NS context the two dynamics are
governed by two coupled differential equations.
This remarkable difference is a consequence of the fact that the
screening current is a dynamical variable itself, and the fluxoid
quantization can be used to eliminate it from the equations in the NS
approach, whereas the quantization rule can be used to eliminate a
phase variable in the XY approach.
A common characteristic of the two sets of equations is that for
$f=0$ a solution is the single junction free running solution. The
difference is that
while for the XY system this is the only solution, the NS system can
allow
for other types of solutions, at least for some parameters values.
The NS model allows the presence of
``propagating solutions'', in the sense that an excitation in a
junction can propagate to the other, giving rise to the so called
``beating solutions'' {\bf \cite{benjacob81,ketoja84}}.
As a consequence the NS approach will reveal a much
richer dynamics than the XY approach; for instance, it can show
hysteresis also in the limit of negligible capacitance (in this limit
the second derivative term disappears), while the XY equation cannot.
Another remarkable difference is that for $f = 1/2$ the XY equation
becomes linear, while the NS set of equations still retains its
nonlinear terms.

To get a deeper insight into the difference between the two approaches
we have numerically evaluated $I_s$ integrating the Eqs. (8,9).
To estimate the importance of $I_s$ in the dynamics we have plotted
the maximum of its absolute value:
%Eq. 10
\begin{equation}
I_s^m = max |I_s/I_0|.
\end{equation}
Initial conditions are always chosen as
$\phi_1=\phi_2=\sin^{-1} \gamma$ if $\gamma < 1$ and zero otherwise,
moreover $\dot{\phi_i}$ was set to zero. We have
used two numerical methods (a simple 4'th order Runge Kutta and a more
elaborate Bulirsch Stoer) obtaining consistent results.
In Fig. 1 we show $I_s^m$ as a function of the bias
current $\gamma$. The two curves are obtained slowly increasing
(diamonds) or decreasing (crosses) the bias current. For both curves
there is a region where the screening current can reach significant
values. To check that $I_s^m$ is really a significant test to
discriminate where solutions of the NS equations are different from
those of the XY model, we have plotted the microscopic dynamics in Fig.
2 for three points, two for zones of the bias in which $I_s^m$ is not
negligible and one for higher bias value. It is evident that in the
first two cases the difference $\phi_2 - \phi_1$ is not just a constant
as assumed by the XY model [see Eq.  (1)], while this is indeed a good
approximation when $I^m_s$ is negligible.

\section{The cell model}
As a first step toward the study of two-dimensional arrays we will
consider
in this section the most elementary cell of a square array
(see Fig. 3).
This cell has already been used to infer the properties of
two-dimensional
arrays (with rf bias) by Sohn and Octavio {\bf \cite{sohn94}}, a row
of such cells has been considered to investigate the stability of
Josephson
array {\bf \cite{filatrella95}}. In the XY approach the equation for
the flux quantization reads (see Fig. 3 for notation):
%Eq. 11
\begin{equation}
V_1 -H_2-V_2+H_1 = 2\pi f
\end{equation}
\noindent
Moreover, for the symmetry of the system also the following identity
holds {\bf \cite{sohn94,filatrella95}}:
%EQ. 12
\begin{equation}
H_1 = - H_2 = H.
\end{equation}
\noindent
The current balance in the four nodes can be written as ($J_X$ denotes
the current flowing through the $RSJ$ elements of the junction $X$):
\begin{eqnarray}
\gamma &=& J_H - J_{V_1}\\
\gamma &=& -J_H - J_{V_2}\\
\gamma &=& -J_H - J_{V_2}\\
\gamma &=& J_H - J_{V_1}
\end{eqnarray}

Taking the sum and difference of the
independent equations the dynamics of the cell in the XY model
is governed by the following two equations:
{\bf \cite{sohn94}}:
%Eq. 17,18
\begin{eqnarray}
\ddot{S} + \alpha \dot{S} + 2\sin\left(\frac{1}{2}S\right)
\cos\left(\frac{1}{2}D\right)&=& -2\gamma \\
\ddot{D} + \alpha \dot{D} + \sin\left(\frac{1}{2}D\right)
\cos\left(\frac{1}{2}S\right) + \sin \left(\frac{1}{2}D -\pi f\right)
&=& 0
\end{eqnarray}
\noindent
where $S=V_1+V_2$ and $D=V_1-V_2$.

For the NS approach, as usual, we have to consider also the
screening current,
while for the same symmetry reasons Eq. (12) still holds. In conclusion
Eq. (11) becomes:
%Eq. 19
\begin{equation}
V_1 -V_2+2H = 2\pi f -\frac{2\pi L I_s}{\Phi_0}
\end{equation}
\noindent
and the equations of motion of the cell are:
%Eq. 20-22
\begin{eqnarray}
\ddot{V_1} + \alpha \dot{V_1} + \sin V_1 &=& - \gamma +
\frac{1}{\beta_l}
\left(V_1-V_2+2H\right)-\frac{2\pi}{\beta_l} f \\
\ddot{V_2} + \alpha \dot{V_2} + \sin V_2 &=& - \gamma -
\frac{1}{\beta_l}
\left(V_1-V_2+2H\right)+\frac{2\pi}{\beta_l} f \\
\ddot{H} + \alpha \dot{H} + \sin H &=& \frac{1}{\beta_l}
\left(V_1-V_2-2H\right)-\frac{2\pi}{\beta_l} f
\end{eqnarray}
\noindent

For comparison we write also Eq.s (20-22) in terms of the variables $S$
and $D$; in this case the Eq.s yield:
\begin{eqnarray}
\ddot{S} + \alpha \dot{S} + 2\sin\left(\frac{1}{2}S\right)
\cos\left(\frac{1}{2}D\right)&
=&
-2\gamma\\
\ddot{D} + \alpha \dot{D} - 2\cos\left(\frac{1}{2}S\right)
\sin\left(\frac{1}{2}D\right)
&=& \frac{2}{\beta_l}(D+2H) - \frac{4 \pi}{\beta_l}f\\
\ddot{H} + \alpha \dot{H} + \sin H &=& \frac{1}{\beta_l}(D+2H)
- \frac{2 \pi}{\beta_l}f
\end{eqnarray}
\noindent

Eq. (23) and Eq. (17) are identical, but Eq. (18) is not easily
recognized
as an approximations of Eq.s (24,25). Therefore
the considerations done for the SQUID apply also here: the two
approaches
differ in the number of equations and it seems difficult to predict
{\it a priori} for which parameter values the screening current is
negligible. Rather, we numerically integrate Eqs. (20-22) and in
Fig. 4 we show the dependence of $I_s^m$ on the parameter $\beta_l$
as a function of the
bias current for two different values of the damping coefficient
$\alpha$;
in Fig. 5 is shown the behavior of the screening current as a function
of
the frustration and the bias current, for the same values of the damping
coefficient. The results shown in these figures can be summarized as
follows:
\begin{itemize}
\item[1)]For $\gamma >> 1$ the screening current decreases. This
corresponds to the observation that (in the overdamped limit) for high
bias current the solutions are nearly sinusoidal (plus a constant slope)
and with a constant phase-shift in
presence of magnetic field {\bf \cite{filatrella95}}. This state
corresponds to negligible screening current.
\item[2)]For high
inductance values (for instance when $\beta_l \ge 2$ for
$\alpha = 0.25$
and $f = 0.5$) the screening current is also negligible. On the
contrary, for the same values of $\alpha$ and $f$ and for $\beta_l$
values down to $0.1$ (a fairly low value for practical arrays
{\bf \cite{note}}), the screening current increases, up to values where
it is not negligible; however for $\beta_l<0.1$ this region shrinks.

\item[3)]The capacitance and the dissipation play an important role in
controlling the screening current effect. Significant screening current
are observed for $\alpha<1$, i.e., for relatively high values of the
capacitance or for low values of the dissipation.
\end{itemize}

It is also important to notice that, as expected, hysteresis
plays an important role. In the figures shown so far we have
always used as initial conditions all the phases equal to zero and at
rest.
As already shown in Fig. 1 a different choice of initial conditions can
lead to a change of the results; This was observed, rather
than changing the initial conditions on the phases, simply sweeping the
current bias starting from values different from zero.
In the cell case simulations shown that there are solutions in the
underdamped limit that
exhibit a significant contribution to the screening current also in
absence of magnetic field.

As we have pointed out in the Introduction, two-dimensional arrays
are of interest
if a large number of junctions can be locked together. Even if we
believe that
the simple cell studied in the previous section offers the distinct
advantage
of simplicity and illustrates the basic mechanism, it is quite natural
to ask if it can also furnish quantitative predictions on larger arrays.
To check that the results obtained so far are not crucially dependent
on the
fact that we are considering only an elementary cell, we have
investigated the
behavior of larger arrays in few cases, using the NS equations
{\bf \cite{nakajima81}}. In Fig. 6 it is shown
the behavior of a $10 \times 10$ array where zones of screening current
comparable with the elementary cell can be clearly seen; in that zones
the overall dynamics can be directly compared with that of the
elementary cell
(see also discussion below). These zones corresponds to a large
range of the
parameters, however, besides these zones in the low damping case
($\alpha=0.25$, see Fig. 6a), a region of very large
maximum screening current is shown: this region corresponds to the
penetration
of static fluxons in the array and does not have any correspondence
in the cell
case; further studies of this region will be carried out in the future.

\section{Microscopic dynamics}
So far we have dealt only with the ``macroscopic'' quantity $I_s^m$.
Although
we claim that this is sufficient to discriminate between states were
model
(1) and model (2) do not give consistent results
(because the set of equations are the
same if $I_s$ is negligible) it is interesting to investigate in which
sense
the dynamics differ when the screening current is not negligible.
For sake
of simplicity we will concentrate on the microscopic dynamics in the
elementary square cell.

We have systematically found that when the screening current
is not negligible
there appear solutions that are close to the well known ``beating
solutions''
for SQUIDs {\bf \cite{benjacob81,ketoja84}}, {\em i.e.}, to
solutions that
correspond to the entry and propagation of fluxons across the cell,
or to the motion of fluxons across the array. To recognize this
we have shown
in Fig. 7 the maximum screening current as a function of the
voltage behavior
together with the IV characteristic. It is evident that a large
screening
current is associated to a resonant step that is reminiscent of the
Fiske steps in long Josephson junctions.
In fact, in the long Josephson junction language resonant structures
are named Fiske step if they occur in presence of magnetic
field, and zero-field steps if they occur in absence of magnetic field
{\bf \cite{barone82}}.

It is also evident that there is a sharp change
in the behavior of $I_s^m$ when the systems switches from the resonant
step to
another solution. To better understand the nature of this transition
we have
plotted in Fig. 8 the voltage across the two vertical junctions for
two values
of the bias current: just before the switch and just after. The
difference
between the two dynamics is that while on the resonant step there is a
large voltage pulse, after the switch the dynamics is much more uniform
with a smaller modulation of the voltage.
We claim that the transition is similar to that observed for
one-dimensional
arrays or for long Josephson junctions {\bf \cite{parment93,ustinov95}}.

A comparison between Fig.s 1,2 and Fig.s 7,8 strongly indicates
that the dynamics are very similar to that observed for the simple
SQUID:
in fact as in the SQUID case resonant states appear in general together
with high values of screening currents and similarly the dynamics
appears to be uniform where screening currents are negligible.

The different nature of the two solutions can be clearly seen
in Fig. 9, where we show a plot of the
time delay across the two vertical junctions of two peaks of voltage.
Indeed, while on the resonant step this time delay roughly tends to a
constant (corresponding to the maximum speed of propagation of a signal
across the system), in the state where the screening currents are
negligible the speed increases with current. This suggests that in the
latter mode there is no actual signal propagating across the cell, but
rather a modulated solution similar to that obtained in Ref. {\bf
\cite{filatrella95}} for overdamped junctions in the high bias limit.
It should be noticed that the very concept of a propagating fluxon is
not applicable to the solution arising
from the XY model:
If we adopt as a working definition of propagating fluxon an excitation
of roughly one flux quantum moving from a cell to another, then to
neglect
screening currents prevents the appearance of such solutions because the
magnetic field is supposed to be uniform across the array.
In this language the XY model does not allow signal
propagation, so we conclude that the XY model can be an approximation
for the
NS equations (as well as any model including screening current
effects through mutual inductances) only if the latter do not carry a
propagating solution.

Since the uniform regime occurs for higher bias,
an interesting question is if the border between those two regions
corresponds to the border between the regions of flux motion and uniform
solutions devised by Lachenmann et al. with the LTSEM (Low Temperature
Scanning Electron Microscopy) technique {\bf
\cite{lachenmann94}}. When comparing the thresholds obtained numerically
with those observed experimentally it is quite arbitrary to decide
how small the screening current should be to lead to the disappearance
of the signal measured by the LTSEM, therefore a detailed quantitative
comparison is not possible on the basis of the results shown here.
However,
the measured threshold for an array $10 \times 10$ critically damped
($\alpha \simeq 1$) is $\gamma \simeq 4$ {\bf \cite{lachenmann94}},
a region where our simulations
predict that the screening current is small: $I_s^m \simeq 0.16$
or roughly $4 \%$ of the bias current at most, i.e., for $f=0.5$.

It is nevertheless possible to notice that both the
screening current and the LTSEM signal indicates an homogeneous
solution for high bias current
{\bf \cite{lachenmann94}}.  If the hypothesis that
the two thresholds are related is true, it is possible to go further
and to speculate that the coherent emission from Josephson
junctions arrays is related to the presence of non negligible
screening currents.

\section{Conclusions}
We have proven that the choice of the more appropriated model
for Josephson
junction arrays depends on the parameters, especially on the
bias current
and the SQUID parameter $\beta_l$. It is possible in fact to show
numerically
that for certain regions of the parameters the contribution to flux
quantization arising from the screening current, neglected in the
so-called
XY model, is important, and this can happen also for relatively small
values of the inductance of the loop. We have traced the origin of
this to
the structure of the equations, noticing that the screening current is
a dynamical variable itself whose value cannot be easily predicted
{\it a priori}, and we were able to investigate its behavior only
numerically. We have also found that the presence of a non
negligible screening current prevents the occurrence of uniform
solutions to set in and rather induces solutions that are known,
in the context of SQUID's, as ``beating solutions''
{\bf \cite{benjacob81,ketoja84}}.
Finally, there are reasons to believe that there might be a
connection between the presence of significant screening
currents and the phenomenon of ``row switching'' observed
with the technique of LTSEM {\bf \cite{lachenmann94}}.

It is worth to recall that we have a) neglected any spread of the
parameters, b) retained self inductances only and neglected mutual
inductances.

\section*{Acknowledgments}
We thank P. Carelli for a critical reading
of this work and T. Doderer, S.G.
Lachenmann, A. Laub, P.L. Christiansen, G. Kalosakas, and
N.F. Pedersen for valuable discussions.  The financial support of ESPRIT
project No. 7100, SCIENCE program No.  SCI*-CT91-0760 ``Coupled
Josephson junctions'', the Human Capital and Mobility grant, program No.
ERBCHRXCT 920068 and of the Progetto Finalizzato ``Superconductive and
Cryogenics Technologies'' is gratefully acknowledged.

\section*{Figure Captions}
\begin{itemize}
\item[1)]Maximum screening current versus bias current for a SQUID.
Parameters of the simulations are: $\alpha = 0.25$, $\beta_l = 1.0$,
$f = 0.5$.
\item[2)]Dynamics of the phase difference of the two
junctions of the SQUID for $\gamma = 0.18$ (solid line), $\gamma =
0.70$ (short dotted line), and $\gamma = 1.0$ (dotted line). Other
parameters are the same as in Fig. 1.
\item[3)]Schematic of the
elementary cell of the two-dimensional square array. Crosses denote the
Josephson elements.
\item[4)]3D plot of the maximum of the screening
current $I_s^m$ versus the bias current $\gamma$ and the SQUID
parameter $\beta_l$ for the elementary cell for (a) $\alpha = 0.25$
and (b) $\alpha = 1.0$.  The frustration is set to $f = 0.5$.
\item[5)]3D plot
of the maximum of the screening current $I_s^m$ versus the bias current
$\gamma$ and the frustration $f$ for the elementary cell for (a)
$\alpha = 0.25$ and (b) $\alpha = 1.0$. The SQUID parameter is set
to $\beta_l = 1.0$.
\item[6)]Maximum screening current as a function
of $\gamma$ and $f$ for an array $10 \times 10$ with $\beta_l=1$.
Parameters of the simulations are: (a) $\alpha = 0.25$,
(b) $\alpha=1.0$.
\item[7)]Voltage (dotted line) and of $I_s^m$ (solid line)
as a function of the bias current in the square cell. Parameters of the
simulations are: $\alpha = 0.25$, $\beta_l = 1$, $f=0.5$.
\item[8)]Dynamics of the voltage across the vertical junctions in the
square cell for
$\gamma = 0.84$ (solid line and dotted line) and $\gamma = 0.86$
(dotted lines with shorter length). Parameters of the simulations
are the same as Fig. 7.
\item[9)]Time delay between two peaks of the voltage in two adjacent
vertical junctions for $\beta_l = 1$ (+) and $\beta_l = 0.5$ (stars).
Parameters of the simulations are: $\alpha = 0.25$, $f=0.5$.
\end{itemize}


\begin{thebibliography}{99}
\bibitem{tilley70}D. R. Tilley, Phys. Lett. {\bf 33}, 205 (1970).
\bibitem{wiesenfeld94}K. Wiesenfeld, S.P. Benz, and P.A.A. Booi,
J. Appl. Phys. {\bf 76}, 3835 (1994).
\bibitem{parment94}R.D. Parmentier, Acta Physica Slovaca {\bf 44}, 303
(1994).
\bibitem{dominguez94}  D. Dom\`inguez, and V. Jos\`e,
Int. J. Mod. Phys. B {\bf 8}, 3749 (1994).
\bibitem{tei}S. Teitel and C. Jayaprakash, Phys. Rev. B {\bf 27},
598 (1983).
\bibitem{hal}T.C. Halsey, Phys.Rev. B {\bf 31}, 5728 (1985).
\bibitem{gro}N. Groenbech-Jensen, A.R. Bishop, F. Falo, and P.S.
Lomdahl, Phys. Rev. B {\bf 45}, 10139 (1992).
\bibitem{nakajima81} K.Nakajima, and Y.Sawada, J. Appl. Phys.,
{\bf 52}, 5732 (1981).
\bibitem{phillips93}J.R. Phillips, H.S.J. van der Zant, J. White,
and T.P.
Orlando, Phys. Rev. B {\bf 47}, 5219 (1993).
\bibitem{reinel94}D. Reinel, W. Dietrich, T. Wolf, and A. Majofer,
Phys. Rev. B {\bf 49}, 9118 (1994).
\bibitem{benz90}S.P. Benz, M.S. Rzchowski, M. Tinkham, and C.J. Lobb,
Phys. Rev. Lett {\bf 64}, 692 (1990).
\bibitem{barone82}A. Barone and G. Patern\`o, {\it Physics and
application of the Josephson effect}, Wiley and sons, NY (1982).
\bibitem{sohn94}L.L. Sohn and M. Octavio, Phys. Rev. B {\bf 49}, 9236
(1994).
\bibitem{tesche77}C.D. Tesche and J. Clarke, J. Low Temp. Phys.
{\bf29}, 301 (1977).
\bibitem{benjacob81} E. Ben-Jacob and Y. Imry, J. Appl. Phys.
{\bf 52} 6806 (1981).
\bibitem{ketoja84}J.A. Ketoja, J. Kurkij\"arvi, and R.K. Ritala,
Phys. Rev. B, {\bf 30}, 3757 (1984).
\bibitem{filatrella95}G. Filatrella, and K. Weisenfield, to
appear on J. Appl. Phys.
\bibitem{note} This limit, as mentioned in the Introduction, refers to
possibility of using practical arrays as coherent sources of
EM-radiation rather than the limit set on the fabrication that is at
about $\beta_l\sim 0.001$ (essentially due to the stray inductance of
the junctions in the array).
\bibitem{parment93}R.D. Parmentier,
``Solitons and Long Josephson Junctionsoo, in {\it The new
Superconducting Electronic}, H.  Weinstock and R.W. Ralsten, eds.
(Kluwer, Dordrecht, 1993), p. 221.
\bibitem{ustinov95}A.V.Ustinov,
M.Cirillo, B.H.Larsen, A.Oboznov, P.Carelli, G.Rotoli, Phys.Rev.B {\bf
51}, 3081 (1995).
\bibitem{lachenmann94}S.G. Lachenmann, T. Doderer,
D. Hoffmann, P.A.A. Booi, and S.P. Benz, Phys. Rev. B {\bf 50}, 3158
(1994).\\ T. Doderer, S.G. Lachenmann, and R.P. Huebener, to appear on
proc. of ``Macroscopic Quantum Phenomena and Coherence in
Superconducting Arrays'', Ed. C. Giovannella, World Scientific.
\end{thebibliography}
\end{document}